\documentclass[aps,pra,epsfigure,twocolumn,nobalancelastpage]{revtex4-2}
\usepackage[colorlinks=true,linkcolor=blue,urlcolor=blue,citecolor=blue,pdfusetitle]{hyperref}
\usepackage[utf8]{inputenc}
\usepackage[english]{babel}
\usepackage{amsmath}
\usepackage{float}
\usepackage{cleveref}
\usepackage[caption = false]{subfig}
\usepackage{blindtext}
\usepackage{amsfonts}
\usepackage[scr=rsfs]{mathalpha}
\usepackage{bbm}
\usepackage{amssymb}
\usepackage{enumerate}
\usepackage{color}
\usepackage{latexsym}
\usepackage{physics}
\usepackage{times,txfonts}
\usepackage{graphicx}
\usepackage{lipsum}
\usepackage{epstopdf}

\begin{document}
\title{Spin-induced multipartite steady-state entanglement of motional modes in hexagonal boron nitride membranes}

\author{Nahid Yazdi}
\affiliation{Department of Physics, Isfahan University of Technology, Isfahan 8415683111, Iran}

\author{Vahid Salari}
\email{vahid.salari1@ucalgary.ca}
\author{Roohollah Ghobadi}
\email{rghobadi@ucalgary.ca}
\affiliation{Institute for Quantum Science and Technology, Department of Physics and Astronomy, University of Calgary, Calgary T2N 1N4, Alberta, Canada}


\begin{abstract}
In this paper, we focus on a scheme in which three high-quality-factor mechanical modes of a hexagonal
boron nitride (hBN) membrane monolayer are coupled to a common optically addressable spin defect
present in the membrane via magnetic field interaction. We show that this coupling induces an effective
phonon-phonon interaction in the dispersive regime and under appropriate magnetic field and microwave
modulation. We derive the Langevin-Heisenberg equations of motion for vibrational modes to analyze
optimal parameter regimes for reaching a physically stable system. We also investigate the effect of
coupling strength on purity and entanglement. Our results demonstrate that bipartite and genuinely
tripartite steady-state entanglement between different vibrational modes of hBN may be achieved in
a broad spectrum of experimental parameters. This study has the potential to enable scalability to
be implemented for the generation of two-dimensional continuous-variable cluster states for universal
quantum computation.
\end{abstract}
\maketitle
\section{Introduction}

Entanglement, a nonclassical correlation of states~\cite{PhysRev.47.777}, is one of the key features of optical, atomic, and electrical quantum systems~\cite{PhysRevLett.109.150502,PhysRevLett.89.253601,pezze2018quantum,hensen2015loophole,cory2004ion} and a powerful physical resource in an effort to exceed classical bounds in quantum computation and information processing. Therefore the characterization of entanglement, especially at a quantitative level, is fundamentally important. One requirement is the ability to entangle multiple particles reliably. Compared with bipartite entanglement, which is now well understood in many aspects, the characterization of multipartite entanglement is still very challenging though a lot of effort has been made~\cite{https://doi.org/10.48550/arxiv.quant-ph/0504163,amico2008entanglement,horodecki2009quantum,briegel2001persistent,pirker2018modular}.

In recent years there has been rapid progress in studying mechanical oscillators as promising candidates for investigating the behavior of macroscopic objects in quantum regimes due to their ability to couple to a variety of degrees of freedom such as photons~\cite{aspelmeyer2014cavity,duan2001long}, superconducting qubits~\cite{lahaye2009nanomechanical,o2010quantum,etaki2008motion}, and single electronic spins~\cite{macquarrie2013mechanical,rabl2010quantum,bennett2013phonon}. Therefore they allow us to establish hybrid mechanical devices. The benefits of such hybrid quantum systems are quite diverse. On the other hand, coupling the resonator to a coherent and fully controllable two-level system (TLS) provides a versatile basis to prepare and detect nonclassical states of mechanical motion and entangle mechanical vibration with a system at the quantum level.

Here, relying on generic spin-phonon coupling, we demonstrate a platform incorporating hexagonal boron nitride (hBN) to create a genuine multipartite steady-state entanglement, which has significant applications in quantum information. Our proposed model has the inherent advantage of providing inspiration to achieve a scalable cluster state on a single chip and could serve as a structure for exploring macroscopic quantum physics and open up new avenues toward universal quantum computation.
hBN flakes have emerged as a compelling wide-band-gap layered material; their hosting of point defects which are single-photon emitters (SPEs)~\cite{tran2016quantum,abdi2018color,PhysRevLett.119.233602} represents significant potential applications in state-of-the-art science and quantum technology. Freestanding two-dimensional (2D) membrane systems, thanks to a number of favorable properties such as achievable extremely low mass density and high elasticity modulus, have considerable capabilities to be promising material playgrounds to explore various basic quantum phenomena in a broad range of fields from quantum computing and quantum communications to quantum sensing.
\section{Model and Hamiltonian}

In fact, hBN has not yet been demonstrated as a spin-mechanics platform experimentally. Here, we use basically the theoretical model proposed in Ref.~\cite{PhysRevLett.119.233602}. The mechanical properties of hBN membranes have been studied before (for instance, see Ref.~\cite{falin2017mechanical}); however, even though the numbers used in this research (and taken from Ref.~\cite{PhysRevLett.119.233602}) may be reasonable, in this respect the mechanical frequencies and quality factor ($Q$) are chosen a bit arbitrarily. Thanks to a high-$Q$ hBN mechanical resonator $(Q \sim 10^{5}$--$10^{6})$ it is within the bounds of possibility to select specific modes. The generic Hamiltonian, $H$, describing our scheme thus is governed by bare spin qubit and mechanical modes and the dynamics of interaction between them~\cite{bravyi2011schrieffer}:

\begin{equation}
\label{H1}
H^{(1)}=-\frac{\Delta}{2}\sigma_{z}+\frac{\Omega_{R}}{2}\sigma_{x}+\sum_{k=1}^{k=3}w_{k}b_{k}^{\dag}b_{k}+\sigma_{z}\sum_{k=1}^{k=3}g_{k}(b_{k}^{\dag}+b_{k}).
\end{equation}

\begin{figure}[t!]
	\centering
	\subfloat[ ]{\includegraphics[scale=0.45]{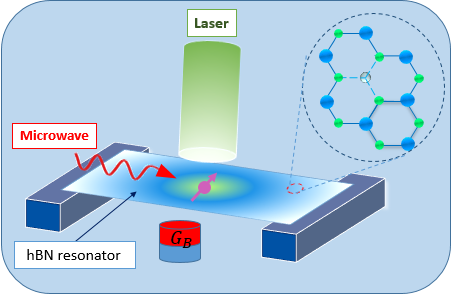}\label{a1}}~
	\subfloat[ ]{\includegraphics[scale=0.53]{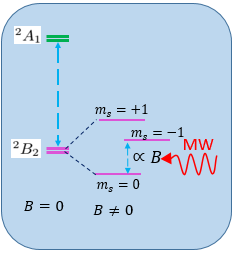}\label{b1}}
	\caption{Schematic description of the proposed setup. \eqref{a1} A single spin can be addressed to couple to three mechanical modes via magnetic coupling.
		\eqref{b1} Level structure of hBN.}
	\label{fig1}
\end{figure}

Here, $\sigma_{x,z}$ refer to Pauli matrices for a two-level system with the convention that $\sigma_{z}=\vert\uparrow\rangle\langle\uparrow\vert-\vert\downarrow\rangle\langle\downarrow\vert$, and $b_{k}(b_{k}^{\dagger})$ denote the bosonic annihilation (creation) operators of the mechanical modes. $\Omega_{R}$ is the Rabi frequency, and $\Delta=\omega_{d}-\omega_{0}$ is the detuning between, for the spin transition $\vert 0\rangle \leftrightarrow \vert-1\rangle$, the tunable spin split-off energy frequency, $\omega_{0}=-g_{s}\mu_{B}B_{z}$, and microwave drive frequency, $\omega_{d}$. Hence $\omega_{0}$ and $\omega_{d}$ provide us with an external independent control over the parameter $\Delta$. $\omega_{k}$ and $a_{0k} =\sqrt{\hbar/ 2m_{k}\omega_{k}}$ are the $k^{th}$ frequency and amplitude of zero-point fluctuations for the megahertz mechanical resonator and effective mass $m$, respectively. $g_{k}=g_{s}\mu_{B} a_{0k}G_{B}$ also characterizes the coupling rate of interaction, which can be controlled by the strength of the magnetic field (Fig.~\ref{fig1}).

\section{Effective~Hamiltonian}
For convenience, we first rotate the spin along the $y$ axis by an angle of $\pi/4$, and tuning the microwave, we gain the capability to take $\Delta=0$ on demand. Now the new Hamiltonian simply reads
\begin{equation}
\label{H2}H^{\prime}=-(\Omega_{R}/2)\sigma_{z}+\sum_{k=1}^{k=3}w_{k}b_{k}^{\dag}b_{k}+\sigma_{x}\sum_{k=1}^{k=3}g_{k}(b_{k}^{\dag}+b_{k}).
\end{equation}

In this frame, the spin has frequency $\Omega_{R}$ and interacts via $H_{\text{int}}=\sigma_{x}\sum_{k=1}^{k=3}g_{k}(b_{k}^{\dag}+b_{k})$ with vibrational modes.

To analyze the underlying physics, we often apply a unitary transformation $U=e^{S}$ to diagonalize our model Hamiltonian. Here, $S$ is the transformation generator and has anti-Hermitian properties.
\begin{equation}
	\label{unitary1}
	 \mathop{H}\limits^{\sim}= U^{\dagger} H^{\prime} U= e^{S}H^{\prime}e^{-S}.
\end{equation}

Using the Baker-Campbell-Hausdorff lemma, we can expand transformation in $S$ as:
 \begin{equation}
  \mathop{H}\limits^{ \sim}=H^{\prime}+[S,H^{\prime}]+\frac{1}{2!}[S,[S,H^{\prime}]]+... =\sum_{n=0}^{\infty}\frac{1}{n!}[S,H^{\prime}]^{(n)}.
 \end{equation}
It is sometimes, however, impractical to exactly diagonalize a Hamiltonian, and so we are concerned with arriving at a low-energy effective Hamiltonian $\mathcal{H}$. To do this, one can perform a Schrieffer-Wolff (SW) transformation~\cite{bravyi2011schrieffer,bukov2016schrieffer} on the Hamiltonian $H$ so that it becomes free from the coupling between $H_{\downarrow}$ and  $H_{\uparrow}$ subspaces. In fact, for the validity of this approach, it is necessary to be in a dispersive regime~\cite{boissonneault2009dispersive,koch2007charge}, i.e., $\delta_{\text{min}}=\underset{k}{\mathrm{\text{min}}}~\vert\delta_{k}\vert=\underset{k}{\mathrm{\text{min}}}\vert\Omega_{R}-\omega_{k}\vert \gg g_{k}~~k=1,2,3,$
where the interaction does not result in
the energy exchange between the spin qubit and motional modes; however, the interaction would demonstrate itself through frequency shift and other physical phenomena as well.
As a result, the qubit remains detuned from all phonons, and consequently, we are in the frame where the qubit degree of freedom decouples from all mechanical modes of the hBN membrane and the transitions between the $H_{\downarrow}$ and  $H_{\uparrow}$ sectors are suppressed. Now we are allowed to project out the high-energy subspace $H_{\uparrow}$, satisfying the condition
 \begin{equation}
 	\label{commutator}
 [H_{0},S]=H_{\text{int}}.
 \end{equation}
In this case $S$ adopts the form $S=\sum_{l=1}^{l=3}g_{l}[\Sigma_{l}(b_{l}^{\dag}\sigma_{+}-b_{l}\sigma_{-})-
\Lambda_{l}(b_{l}^{\dag}\sigma_{-}-b_{l}\sigma_{+})]$, where $\Sigma_{l}=(\Omega_{R}+w_{l})^{-1}$ and $\Lambda_{l} =({\Omega_{R}-w_{l}})^{-1}$ (see Appendix~\ref{ApB}).
Following this expansion up to the third order of the interaction
 $ \mathop{H}\limits^{ \sim}=H_{0}+\frac{1}{2}[S,H_{\text{int}}]+O(H_{\text{int}}^{3})$ and neglecting the constant term, the effective Hamiltonian is obtained as
\begin{equation}
\label{H}\mathop{H}\limits^{ \sim}=\frac{\Omega_{R}}{2}\sigma_{z}+\sum_{k=1}^{3}w_{k}b_{k}^{\dagger}b_{k}+
\Omega_{R}\sigma_{z}\sum_{k,l=1}^{3}g_{l}g_{k}J_{l}
(b_{k}^{\dagger}+b_{k})(b_{l}^{\dagger}+b_{l}),
\end{equation}
where $J_{l}=\Sigma_{l}\Lambda_{l}=(\Omega_{R}^{2}-w_{l}^{2})^{-1}$.
Since we focus on the achievable situation of a large detuned regime~\cite{PhysRevLett.119.233602,leroux2018enhancing,ashhab2010qubit,casanova2010deep}, the spin tends to be trapped
corresponding to mostly remaining in the ground electronic state (see Appendix~\ref{ApA}), so that $\mathcal{H}=\mathop{H}\limits^{ \sim}\otimes\vert \downarrow\rangle\langle\downarrow \vert$~\cite{hwang2015quantum,hwang2016quantum}. By adopting the adiabatic approximation, the behavior of the
mechanical mode becomes more apparent, so that the resulting effective Hamiltonian, $H_{eff}=\langle \downarrow\vert \mathcal{H}\vert\downarrow \rangle$, contains only bosonic operators $b_{k}$ and is quadratic in terms of them. Hence we reach our goal of having a Gaussian Hamiltonian. Appearing terms such as $b_{l}^{\dag}b_{k}^{\dag}$ and $b_{k}b_{l}$ can be considered as an intuitive indication as to why the mechanical modes are entangled, and $e^{i\theta G_{kl}(b_{k}^{\dagger}+b_{k})(b_{l}^{\dagger}+b_{l})}$ acts as a controlled-Z gate to create entanglement between modes $k$ and $l$, where $\theta=2\sqrt{2}\Omega_{R}\langle \downarrow \vert\sigma_{z} \vert\downarrow\rangle$ and $G_{kl}=J_{l}g_{l}g_{k}$.

The model presented in this paper works in the dispersive regime in which one can adiabatically eliminate the spin degree of freedom and one has an effective Schrieffer-Wolff treatment, ending up in an effective Hamiltonian of three coupled oscillators. Therefore the paper in practice studies entanglement in a simple system of three linearly coupled oscillators.
\section{Quantum~Langevin-Heisenberg~Equations}
 We study the dynamics of our system by adopting a quantum Langevin equation (QLE) treatment~\cite{wang2013reservoir,vitali2007optomechanical}. For a detailed analysis of the problem, we take into account the mechanical damping of the membrane.
The Hamiltonian $H_{eff}$ gives rise to the equations of motion
\begin{subequations}
	\begin{align}
		\label{eqmotionx}\dot{x}_{m}&=-\gamma_{m}x_m+w_m p_m+\sqrt{2\gamma_{m}}x_{m}^{in}\\
		\label{eqmotionp}\dot{p}_{m}&=-w_{m}x_{m}-\gamma_{m}p_{m}
		+\alpha\sum_{k=1}^{k=3}g_{m}g_{k}x_{k}(J_{k}+J_{m})+\sqrt{2\gamma_{m}}p_{m}^{in}.
	\end{align}
\end{subequations}
Here $\gamma_{m}$ is the mechanical damping rate for each mode, and $x_{m}=(b_{m}+b_{m}^{\dagger})/\sqrt{2}$ and $p_{m}=i(b_{m}^{\dagger}-b_{m})/\sqrt{2}$ are the dimensionless position and momentum quadratures of the mechanical modes, $[x_{m},p_{m}]=i$, where the input thermal noise of the mechanical modes $x_{m}^{in} , p_{m}^{in}$, with zero mean values, satisfies the commutation relation
$[\hat{\mathscr{O}}(t),\hat{\mathscr{O}}^{\dagger}(t^\prime)] =\delta(t-t^\prime)$.
In addition, in the limit of high mechanical quality factor $(Q_{m}\gg 1)$, which is required for
emerging quantum effects such as entanglement, the membrane thermal noise $b_{m}^{in}$ can be faithfully considered as a Markovian noise whose nonzero correlation functions are given by $\langle \hat{\mathscr{O}}(t)\hat{\mathscr{O}}^{\dagger}(t^\prime)+\hat{\mathscr{O}}^{\dagger}(t^\prime)\hat{\mathscr{O}}(t)\rangle/2=\gamma_{m}(1+2\bar{n}_{m})\delta(t-t^\prime)$,
where $(\hat{\mathscr{O}}=x_{m}^{in} , p_{m}^{in})$ and $\bar{n}_{m}=[exp(\hbar\omega_{m}/k_{B}T)-1]^{-1}$
is the mean number of thermal phonons in the absence of the spin-motion coupling, with $k_{B}$ and $T$ being the Boltzmann constant and the temperature of the mechanical bath, respectively~\cite{law1995interaction,gardiner2004quantum}. The compact form of the QLEs can be written as
\begin{equation}
\dot{\mathcal{R}}(t)=\mathcal{A}\mathcal{R}(t)+\mathcal{N}(t)\label{difR}
\end{equation}
where we define the vector $\mathcal{R}^{T}(t) =(x_{1},p_{1},x_{2},p_{2},x_{3},p_{3})^{T}$; the noise vector $\mathcal{N}_{i}=\sqrt{2\xi_{i}}\mathcal{R}^{in}_{i}$, where $\xi^{T}=(\gamma_{1},\gamma_{1},\gamma_{2},\gamma_{2},\gamma_{3},\gamma_{3})^{T}$; and the kernel $\mathcal{A}$ as
\begin{widetext}
	\begin{align*}
	\mathcal{A} & =\begin{pmatrix}
	-\gamma_{1}&  w_{1} & 0& 0&0&0\\
	-w_1+G_{1,1}&  -\gamma_{1} & G_{1,2}& 0&G_{1,3}&0\\
	0&  0 & -\gamma_{2}& w_{2}&0&0\\
	G_{1,2}&  0 & -w_{2}+G_{2,2}& -\gamma_{2}&0&G_{2,3}\\
	0& 0 & 0& 0&-\gamma_{3}& w_{3}\\
	G_{3,1}&  0 & G_{3,2}& 0&-w_{3}+G_{3,3}&-\gamma_{3}
	\end{pmatrix}
	\end{align*}
	\vspace*{1pt}
	where $G_{i,j}=\theta g_{i}g_{j}(J_{i}+J_{j})$.
\end{widetext}
It is worth mentioning that having negative real parts of the eigenvalues of kernel $\mathcal{A}$ is essential to guarantee the existence of steady-state entanglement properties. Henceforth, we will focus on the parameter regime that satisfies this and ensure that the steady state of the system is stable. These conditions can be investigated analytically by applying the Routh-Hurwitz criterion~\cite{grandshteyn1980table}, which maybe is cumbersome. One can show that the zero-mean Gaussian steady-state covariance matrix (CM) $V$, with components $V_{i,j}=\langle \mathcal{R}_{i}(\infty)\mathcal{R}_{j}(\infty) +\mathcal{R}_{j}(\infty)\mathcal{R}_{i}(\infty)\rangle/2$, which is fully characterized by the $6\times 6$ stationary correlations~\cite{braunstein2005quantum}, fulfills the Lyapunov equation:
\begin{equation}
\mathcal{A}V+ V\mathcal{A}^{T} =-D,
\end{equation}
which is a linear equation for $V$ and can be straightforwardly solved, where $D_{ij}=\delta_{ij}\xi_{i}(2\bar{n}_{i}+1)$.
\section{Steady-State Entanglement}
\begin{figure}[h!]
	\centering
	\subfloat[ ]{\includegraphics[scale=0.36]{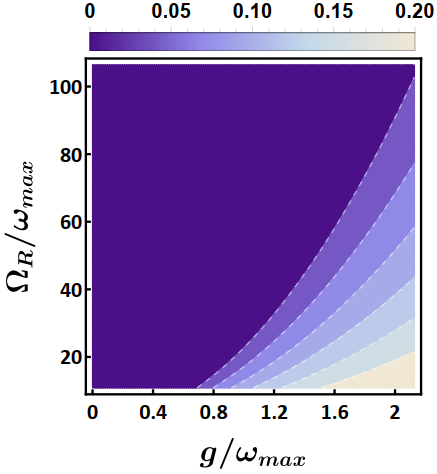}\label{a2}}
	~\subfloat[ ]{\includegraphics[scale=0.36]{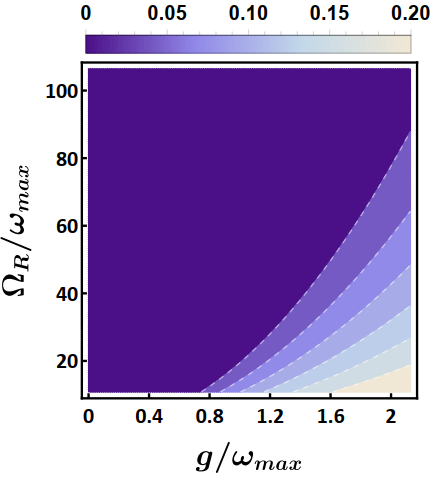}\label{b2}}\\
	~\subfloat[ ]{\includegraphics[scale=0.36]{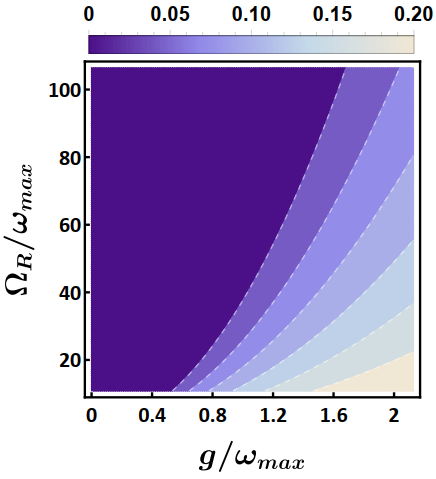}\label{c2}}
	~\subfloat[ ]{\includegraphics[scale=0.36]{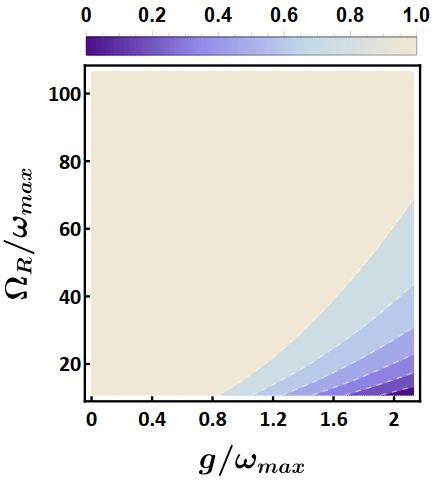}\label{d2}}
	\caption{Density plot of the steady-state bipartite entanglement (logarithmic negativity $E_{\mathscr{N}}$) for \eqref{a2} $E_{\mathscr{N}} ^{12}$, \eqref{b2} $E_{\mathscr{N}} ^{13}$, and \eqref{c2} $E_{\mathscr{N}}^{23}$, and \eqref{d2} density plot of the stability of the system, vs the normalized Rabi frequency of the classical driving field $\Omega_{R}/\omega_{1}$ and the ratio of coupling constant and mode frequency $g/\omega_{1}$ for the parameters given in the text.}
 \label{fig2}
\end{figure}

The covariance $V$ encompasses all the information we need to compute the steady-state entanglement~\cite{adesso2006multipartite}. As an entanglement measure, we introduce the logarithmic negativity $E_{\mathscr{N}}$, which can be used to
establish the conditions under which the mechanical modes are entangled, and for any bipartite Gaussian state it is defined as~\cite{adesso2007entanglement}
\begin{equation}
\label{E}E_{\mathscr{N}}=\text{max}[0,-\text{ln}~(2\nu^{-})],
\end{equation}
where $\nu^{-}\equiv 2^{-1/2}[\Lambda(\sigma)-\sqrt{\Lambda(\sigma)^{2}-4~\text{Det}~\mathcal{Y}}]^{1/2}$ is the smallest symplectic eigenvalue of the partially transposed bipartite CM $\tilde{\mathcal{Y}}=\mathcal{P}^{1\vert2}  \mathcal{Y}~ \mathcal{P}^{1\vert2}$,
where $\mathcal{Y}$ is the $4\times 4$ reduced covariance matrix that contributes to the two mechanical modes of interest, obtained by removing
in $V$ the rows and columns of the uninteresting mode, and
$P^{1|2} = diag(1,-1, 1, 1)$ is the matrix that realizes partial transposition at the level of CMs~\cite{PhysRevLett.84.2726}.
\begin{align*}
\label{Y}
\mathcal{Y} & =\begin{pmatrix}
&\mathcal{B} & \mathcal{C} \\
& \mathcal{C}^{T} & \mathcal{D}\\
\end{pmatrix}
\end{align*}
and $\Lambda(\sigma)\equiv \text{Det}~\mathcal{B}+\text{Det}~\mathcal{D}-2\text{Det}~\mathcal{C}$, where $\mathcal{C}$ corresponds to the correlations between these two modes.
We find the optimal effective interaction regime with the aim of maximizing the generated tripartite entanglement among three membrane oscillation modes~\cite{gardiner2004quantum}. Figures~\ref{a2}--\ref{c2} show the behavior of the degree of bipartite entanglement extract from Eq.~\eqref{E}. These results are found for mechanical modes that are characterized by different, but quite close, frequencies, with $\omega_{1}/2\pi= 2.70$ MHz, $\omega_{2}/2\pi=3.80$ MHz, and $\omega_{3}/2\pi=4.70$ MHz, in the case of $a_{0k} \thickapprox 5\times10^{-13}$ m and $G_{B}\thickapprox 270$ G/nm at the position of the spin qubit. In this case, we focus on the regime of low temperature, $T=0.05$ mK, which is still hardly achievable in cryogenics, and one may need adiabatic demagnetization while probably this would not be enough. Based on the current model, entanglement may not survive at higher temperatures mainly due to the small coupling inherent in the scheme, and therefore the model should be improved to be applicable for higher temperatures to be testable in the laboratory. So, it is necessary to cool
down mechanical modes to their ground states, i.e., $\bar{n}_{k}\ll 1$ for $k=1,2,3$. Therefore, under appropriate conditions, as is demonstrated in Fig.~\ref{fig2}, one can achieve a significant entanglement of up to $E_{\mathscr{N}}=0.2$. It is also interesting to note that in all the above cases, even in a weak-interaction regime ($\omega_{k}<g_{k}$, where $k=1,2,3$), entanglement emerges.
Since there are no neat analytical relations between parameters when the system enters the unstable region, it is convenient to find this area by numerical solution. No instability has been found lying in this accessible parameter regime that we have employed. We also investigate the global purity of our system since it is relevant for experiments in quantum foundations and in quantum information processing. For an $n$-mode Gaussian state the purity is simply evaluated by $\mu=1/(2^{n}\sqrt{\text{Det}~V})$~\cite{adesso2004determination}.

Figure~\ref{d2} reveals that the fact that the steady state is highly entangled does not necessarily imply that the steady state is highly pure. Indeed, if one increases the coupling strength, then the steady state is not only highly entangled but also highly impure.
\begin{figure}[t!]
	\centering
	\subfloat[ ]{\includegraphics[scale=0.29]{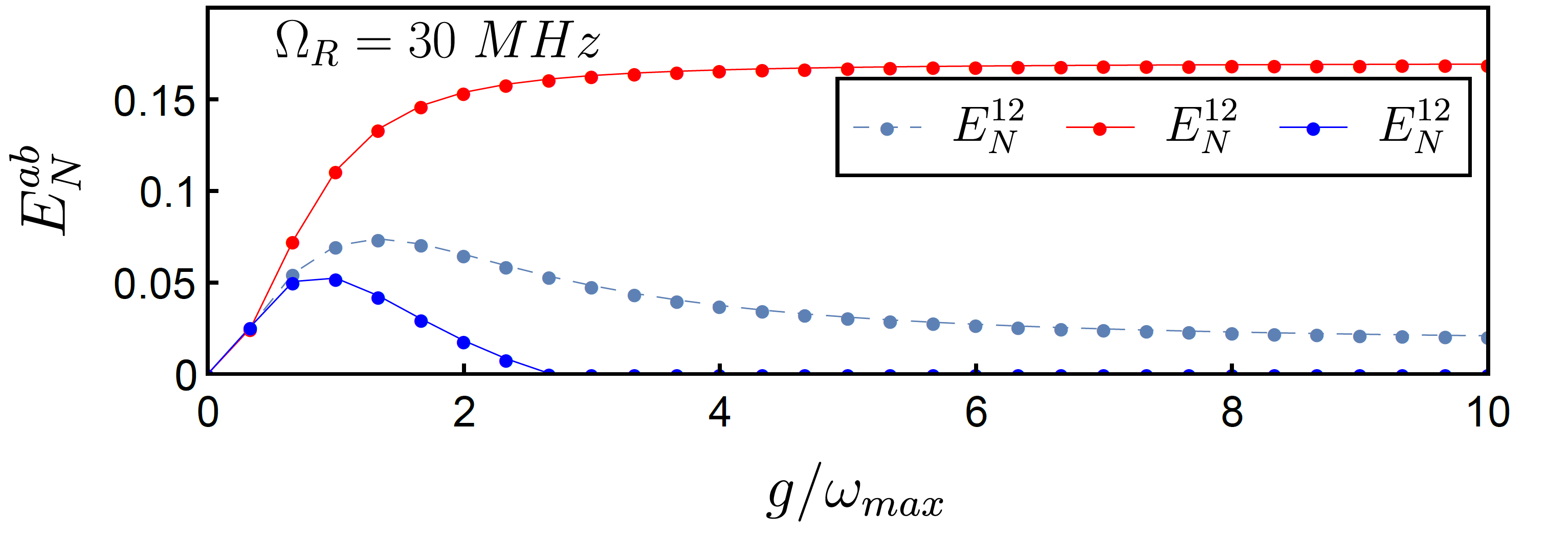}\label{a3}}\\
	\subfloat[ ]{\includegraphics[scale=0.455]{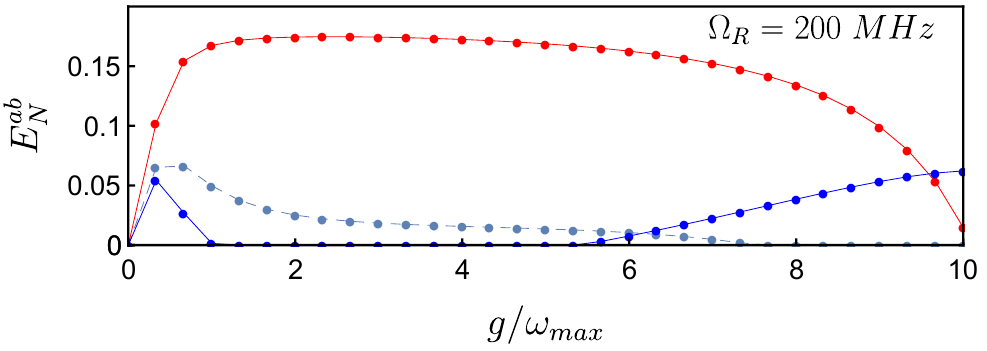}\label{b3}}
	\caption{Plot of logarithmic negativity $E_{\mathscr{N}}$ of the three bipartite subsystems $E_{\mathscr{N}}^{12}$ (phonons 1 and 2), $E_{\mathscr{N}}^{23}$ (phonons 2 and 3), and $E_{\mathscr{N}}^{13}$ (phonons 1 and 3)
	vs the normalized coupling constant $g/\omega_{\text{max}}$ in which $\omega_{\text{max}}=\omega_{3}$ and $g=g_{1}$ for two different values of the Rabi frequency, \eqref{a3} $\Omega_{R}/2\pi=30$ MHz and \eqref{b3} $\Omega_{R}/2\pi=200$ MHz, for $\bar{n}_{k}=0$ $(k=1,2,3)$.}
	\label{fig3}
\end{figure}
In Figs.~\ref{a3}--\ref{b3} we survey the potential of the classical drive with two different values of $\Omega_{R}$ as a controllable tool to determine the logarithmic negativity of the three bipartite phonon subsystems which have frequencies $\omega_{1}/2\pi=1$ MHz, $\omega_{2}/2\pi=7$ MHz, and $\omega_{3}/2\pi=9$ MHz when $Q =1\times 10^{6}$. We set $g_{2} =1.2g_{1}$ and $g_{3} =1.5g_{1}$. We point out that the maximum of the phonon-phonon entanglement does not necessarily appear in the strong- or deep-strong-coupling regime, and even the entanglement between one of the two modes may be completely suppressed, while in the weak-coupling regime, by increasing the coupling strength, all bipartite entanglements tend to rise monotonically. Therefore there is not a direct relation between coupling and entanglement, and the response of the system is strongly dependent on the value of $\Omega_{R}$.


In Fig.~\ref{b3} it is obvious that enhancement of the purity is possible by increasing the Rabi frequency of the microwave drive.

\section{Genuine Tripartite Entanglement}

It would be useful to determine the entanglement class of the system state by applying the results of Refs.~\cite{adesso2006multipartite,adesso2007entanglement}, which have provided a necessary and sufficient criterion for the determination of the class in the case of tripartite CV Gaussian states; this criterion is directly computable. This classification criterion is mostly based on the nonpositive partial transposition (NPT) criterion proved in~\cite{adesso2007entanglement}, which is necessary and sufficient for 1 $\times$ N bipartite CV Gaussian states. This is also confirmed by the fact that such a state is a fully inseparable tripartite CV entangled state in the parameter regime of Fig.~\ref{fig2}, for a wide range of parameter regimes.
To check that genuine tripartite entanglement is shared between the subsystems, we utilize the minimum residual continuous-variable tangle (contangle) $E^{\text{min}}_{\tau}$ as a $bona\;\; fide$ quantifier~\cite{adesso2007entanglement,PhysRevLett.84.2726,duan2000inseparability}
\begin{equation}
\label{S}
E^{\text{min}}_{\tau}\equiv \underset{(i,j,k)}{\mathrm{\text{min}}} [E^{i\vert(ij)}_{\tau}- E^{i\vert j}_{\tau}-E^{i\vert k}_{\tau}]~~(i,j,k=1,2,3),
\end{equation}
which is invariant under all permutation of the symbols $(i,j,k)$ (for more details see Appendix~\ref{ApC}).
\begin{figure}[H]
	\centering
	\subfloat[ ]{\includegraphics[scale=0.4]{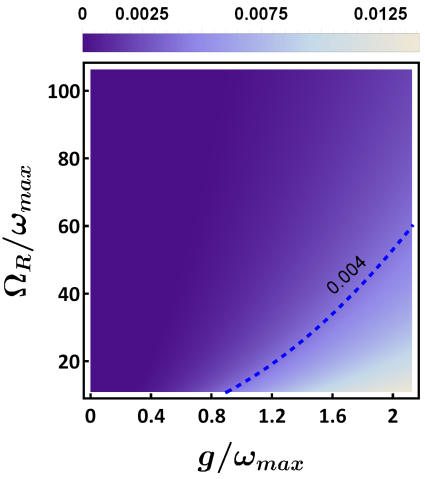}\label{a4}}
	\subfloat[ ]{\includegraphics[scale=0.4]{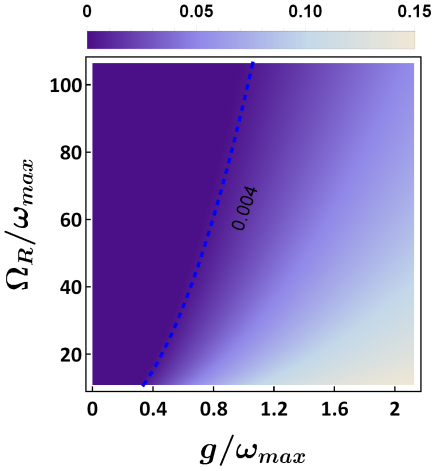}\label{b4}}
	\caption{\eqref{a4} Genuine tripartite entanglement as measured by the minimum residual contangle $E^{\text{min}}_{\tau}$. \eqref{b4} Bipartite entanglement for $\omega_{1}/2\pi=\omega_{2}/2\pi=\omega_{3}/2\pi=4.70$ MHz and $g_{1}=g_{2}=g_{3}$. The other parameters are the same as in Fig.~\ref{fig2}.}
	\label{fig4}
\end{figure}
\vspace*{-3mm}
\section{Conclusions}
The analysis in this paper mainly relies on bipartite and tripartite entanglement in a set of linearly coupled oscillators, which has been studied extensively within cavity optomechanics, even though often involving hybrid bosonic systems (e.g., see Refs.~\cite{PhysRevA.77.050307,Genes_2008,genes2009quantum} and, more recently, Ref.~\cite{PhysRevResearch.4.033112}), and we have used an identical analysis for a hBN layer here, too. However, in fact, the above systems are not quite similar, and the models may need to be generalized and compared. In conclusion, based on the scheme introduced here, the feasibility of
generating tripartite steady-state entanglement between
different vibrational modes of a 2D hBN membrane (which is freestanding and subjected to a strong gradient magnetic field), where three phonons are simultaneously coupled to a common qubit, is examined numerically by calculating the logarithmic negativity. The scheme only requires driving by a classical microwave light, working in very low temperatures up to a few millikelvins. Therefore the created entanglement is extremely fragile in regard to temperature, and so the overwhelming environmental heating has a significant impact. We have shown that with experimentally reachable parameters, the steady state of membrane phonons can be a bipartite CV entanglement for all the possible bipartite subsystems thanks to the mediating action of the spin. In particular, the stationary state of the system exhibits genuine tripartite entanglement. For further study it would be very interesting to demonstrate the scalability of our approach, so that one can analyze this system when more than three phonons have been provoked to build networks of multiple mechanical modes.
The results obtained in this paper are only valid at very low temperatures close to absolute zero. Therefore, in order to have a real application, the entanglement should be checked for higher temperatures with the drive of mechanical modes in a next research prospect; then this approach has potential for applications in the field of quantum information, and one could promote its extension to promising platforms for the study of macroscopic quantum phenomena, with numerous developments in quantum sensing as well as one-way quantum computation.

\appendix

\section{Adiabatic Elimination of the Spin Degree of Freedom}\label{ApA}

The main reason for using the covariance matrix approach in this paper is the large dimensions of the system's Hilbert space. This system consists of three phonon modes (continuous variable) and a spin; the solution of the master equation in the steady state by software such as QUTIP and MATHEMATICA is practically very difficult due to limited random access memory (RAM). In the case of adding the Liouvillian part (spin noise effects) to the covariance matrix, because this approach applies to continuous variables, due to the limited spin Hilbert space, it is impossible to enter the spin effect in this covariance matrix. Therefore it is necessary to work in a regime that can eliminate the spin degree of freedom adiabatically (see Refs.~\cite{PhysRevLett.119.233602,PhysRevA.75.032302,PhysRevA.91.043801} the Supplemental Material of Ref.~\cite{hwang2016quantum}).
\begin{figure}[H]
	\centering
	\subfloat[ ]{\includegraphics[scale=0.230]{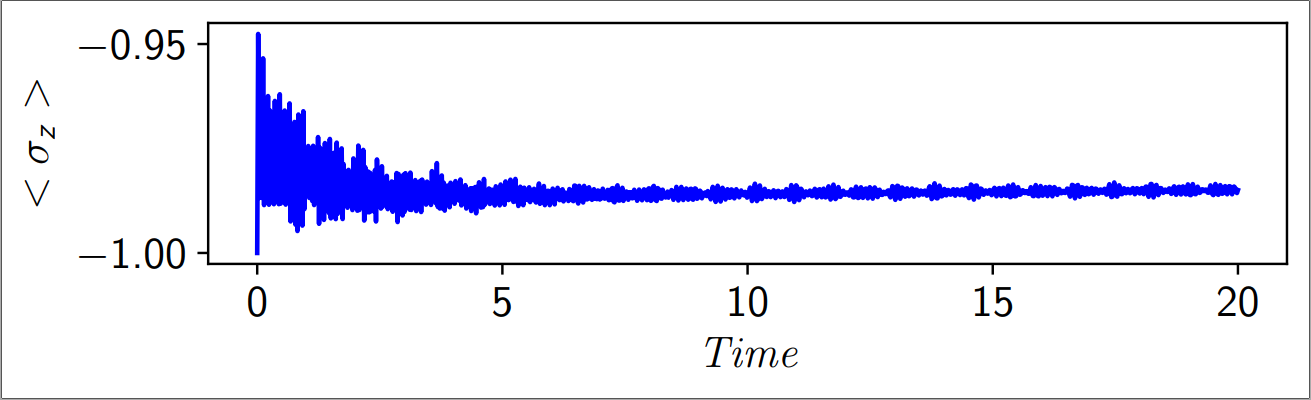}\label{a5}}
	\caption{Expectation value for the operator $\sigma_{z}$ as time goes by, in the far-detuning regime. The other parameters are the same as in the main text.}
	\label{fig5}
\end{figure}
The parametric regime in this paper is essentially the same as the parametric regime reported in~\cite{PhysRevLett.119.233602}.
In this paper, the spin degree of freedom has been removed from the mechanical mode based on an adiabatic approximation to create a squeezed state. According to the available references~\cite{boissonneault2009dispersive, koch2007charge, PhysRevA.91.043801, PhysRevA.75.032302, PhysRevLett.117.015502, Schmidt_2012}, this approximation works well in the dispersive regime. Thanks to large detuning between the frequency of the mechanical modes and the spin transition frequency there is virtually no effective energy exchange between them, and so the spin remains in the ground state (see Fig.~\ref{fig5}).

\section{Analytical Solution for the System Dynamics}\label{ApB}
The most crucial step in performing SW transformation is to get the generator of the
transformation. Once the generator is calculated, the rest of the calculation is quite straightforward.
In this Appendix, we present an explicit method
to compute the generator of the SW transformation for the original Hamiltonian~\eqref{H}:

\begin{equation}
H^{\prime}=H_{0}+H_{int}.
\end{equation}

$H_0$ is the diagonal part and $H_{int}$ is the off-diagonal
part of the full Hamiltonian $H^{\prime}$.
Considering the condition~\eqref{commutator}, it is obvious that the most general form of $S$ will be as follows:
\begin{equation}
S=\vec{\alpha}.\vec{\sigma}\sum_{l=1}^{l=3}f_{l}(b_{l}^{\dag}+b_{l})+\vec{\beta}.\vec{\sigma}\sum_{l=1}^{l=3}h_{l}(b_{l}^{\dag}-b_{l}),
\end{equation}
where $\vec{\sigma}=(\sigma_{0},\sigma_{1},\sigma_{2},\sigma_{3})\equiv(I,\sigma_{x},\sigma_{y},\sigma_{z})$ and $\vec{\alpha}$, $\vec{\beta}$, $f_{l},~h_{l}$ are the undefined coefficients of $S$.
Now by applying the condition that the generator is anti-Hermitian, i.e., $S^{\dag}=-S$, one option can be
\begin{subequations}
	\begin{align}
	\label{coeff1}
\vec{{\alpha}}=&-\vec{\alpha}^*,\\
	\label{coeff2}
	h_{i}=&h_{i}^*,\\
	\label{coeff3}
	f_{i}=&f_{i}^*;
	\end{align}
\end{subequations}
then we can replace $\alpha_{i} \to -i\alpha_{i}$, where the new $\alpha$ is real.

Now we will use the condition~\eqref{commutator} to completely determine the rest of the coefficients.

\begin{align}
\label{A4}[H_{0},S]=[\frac{\Delta}{2}\sigma_{z}+\sum_{k=1}^{k=3}w_{k}b_{k}^{\dag}b_{k}, S]=\sigma_{x}\sum_{l=k}^{k=3}g_{k}(b_{k}^{\dag}+b_{k})\notag\\
=\sum_{l=1}^{l=3}(b_{l}^{\dag}+b_{l})[\sum_{s=0}^{s=1}\sum_{m=0}^{m=3}-\Delta f_{l}\alpha_{s}\left(\varepsilon_{szm}\sigma_{m}+w_{l}h_{l}\beta_{s}\sigma_{s}\right)]\nonumber\\
+\sum_{l=1}^{l=3}(b_{l}-b_{l}^{\dag})
[\sum_{s=0}^{s=1}\sum_{p=0}^{p=3}i\Delta h_{l}\left(\beta_{s}\varepsilon_{szp}\sigma_{p}+iw_{l}f_{l}\alpha_{s}\sigma_{s}\right)]\nonumber\\
\end{align}

We note that $(b_{k}+b_{k}^{\dag})$, $(b_{k}-b_{k}^{\dag})$, and the components of $\vec{\sigma}$ are linearly independent.
According to this remark and by comparing the two sides of Eq.~\eqref{A4}, the coefficient of $(b_{l}-b_{l}^{\dag})$ must be zero, and the coefficient of $(b_{l}^{\dag}+b_{l})$ must be equal to $-g_{l}\sigma_{x}$.

\begin{subequations}
	\begin{align}
	\label{con1}
&\sigma_{s}=\sigma_{m}=\sigma_{p}~~\text{and}~~\alpha_{y}\neq 0~~~~ \beta_{x}\neq 0, \\
	\label{con2}
&\sum_{l=1}^{l=3}\sum_{s=0}^{s=3}\Delta f_{l}\alpha_{s}\sum_{m=0}^{m=3}\left(\varepsilon_{szm}\sigma_{m}+w_{l}h_{l}\beta_{s}\sigma_{s}\right)=
\sum_{l=1}^{l=3}g_{l}\sigma_{x}.
	\end{align}
\end{subequations}
So we arrive at

\begin{subequations}
	\begin{align}
	\label{con3}
	& \Delta f_{l}\alpha_{y}-w_{l}h_{l}\beta_{x}=g_{l}, \\
	\label{con4}
	&i(\Delta h_{l}\beta_{x}\varepsilon_{xzy}+w_{l}f_{l}\alpha_{y})\sigma_{y}=0.
	\end{align}
\end{subequations}

$\alpha_{y}, \beta_{x}$ are free variables; so we let them be equal to $2$. Solving for $f_{l}$ and $h_{l}$, we obtain
\begin{align}
f_{l}=\dfrac{\Delta }{\Delta^{2}-w_{l}^{2}}g_{l} ,\nonumber\\
h_{l}=\dfrac{w_{l} }{\Delta^{2}-w_{l}^{2}}g_{l}.
\end{align}
We rewrite $S$:
\begin{align}
S=\sum_{l=1}^{l=3}\dfrac{ g_{l}}{\Delta+w_{l}}(b_{l}^{\dag}\sigma_{+}-b_{l}\sigma_{-})-
\dfrac{ g_{l}}{\Delta-w_{l}}(b_{l}^{\dag}\sigma_{-}-b_{l}\sigma_{+}).
\end{align}

\section{Quantification of Gaussian Tripartite Entanglement}\label{ApC}

For the study of tripartite entanglement, we adopt a quantitative
measure, the residual contangle $
 E^{i\vert(jk)}_{\tau}= E^{i\vert(ij)}_{\tau}- E^{i\vert j}_{\tau}-E^{i\vert k}_{\tau}~~(i,j,k=1,2,3)$~\cite{adesso2007entanglement,2006NJPh....8...15A}, where $E^{p\vert q}_{\tau}$ is the contangle of subsystems of $p$ and $q$ ($q$
 contains one or two modes), which is a proper entanglement
 monotone defined as the squared logarithmic negativity
 \cite{adesso2007entanglement}. To calculate the $one\text{-}mode\text{-}vs\text{-}two\text{-}modes$ logarithmic negativity $E^{i\vert(jk)}$, one only needs to follow the definition of Eq.~\eqref{E} simply by replacing $\tilde{\mathcal{Y}}=\mathcal{P}^{1\vert2}  \mathcal{Y}~ \mathcal{P}^{1\vert2}$ with  $\mathcal{Y}=\mathcal{P}^{i\vert jk}  \mathcal{Y}~ \mathcal{P}^{i\vert jk}$, where $\mathcal{P}^{1\vert 23}=diag(1,-1,1,1,1,1)$, $\mathcal{P}^{2\vert 13}=diag(1,1,1,-1,1,1)$, and $\mathcal{P}^{3\vert 12}=diag(1,1,1,1,1,-1)$ are partial transposition matrices. The residual contangle satisfies the monogamy of
 quantum entanglement $ E^{i\vert(jk)}_{\tau}\geq0$, i.e.,
 \begin{equation}
 \label{con}
 E^{i\vert jk}_{\tau}\geq E^{i\vert j}_{\tau} + E^{i\vert k}_{\tau}.
 \end{equation}
 A bona fide quantification of CV tripartite entanglement is
 provided by the minimum residual contangle~\cite{adesso2007entanglement,2006NJPh....8...15A}
\begin{equation}
\label{contangle}
E^{\text{min}}_{\tau}\equiv \underset{(i,j,k)}{\mathrm{\text{min}}} [E^{i\vert(ij)}_{\tau}- E^{i\vert j}_{\tau}-E^{i\vert k}_{\tau}]~~(i,j,k=1,2,3),
\end{equation}
which ensures that $E^{\text{min}}_{\tau}$ is invariant under all permutations of the modes and is thus a genuine three-way property of any
three-mode Gaussian state.

\bibliography{MainRefs.bib}
\end{document}